\documentclass[twocolumn,superscriptaddress,nobibnotes,footinbib,amsmath,floatfix,aps,prl,citeautoscript,reprint]{revtex4-1}

\usepackage{graphicx}
\usepackage{dcolumn}
\usepackage{bm}
\usepackage{float}
\usepackage{color}
\usepackage[draft]{changes} 
\usepackage[normalem]{ulem}
\definechangesauthor[name={yi}, color=blue]{Yi}

\newcommand{\nw}{Department of Materials Science and Engineering, Northwestern University, Evanston, IL 60208, USA}
\newcommand{\lbnl}{Energy Technologies Area, Lawrence Berkeley National Laboratory, Berkeley, CA 94720, USA}

\newcommand{\yale}{Department of Applied Physics, Yale University, New Haven, CT 06520, USA}
\newcommand{\esi}{Energy Sciences Institute, Yale University, West Haven, CT 06516, USA}

\usepackage{graphicx}
\usepackage{sidecap}
\usepackage{tabularx}
\usepackage{sidecap}
\newcommand{\etal}{\textit{et al}. }

\begin{document}

\title{Leveraging Electron-Phonon Interaction to Enhance Thermoelectric \\ Power Factor in Graphene-Like Semimetals}

\author{Yi Xia}
\email{yimaverickxia@gmail.com}
\affiliation{\nw}

\author{Junsoo Park}
\affiliation{\lbnl}

\author{Vidvuds Ozoli\c{n}\v{s}}
\affiliation{\yale}
\affiliation{\esi}

\author{Chris Wolverton}
\email{c-wolverton@northwestern.edu}
\affiliation{\nw}

\date{\today}

\begin{abstract}

Electron-phonon interactions (EPI) are presumably detrimental for thermoelectric performance in semiconductors because it limits carrier mobility. Here we show that enhanced EPI with strong energy dependence offers an intrinsic pathway to significant increase in the Seebeck coefficient and the thermoelectric power factor, particularly in the context of two-dimensional (2D) graphene-like Dirac bands. The increase is realized by enabling electron energy filtering through preferential scattering of electron/hole carriers. We prove this concept by implementing first-principles computational methods with explicit treatment of EPI for a 2D gapless MoS$_{2}$ allotrope, which has both massless Dirac bands and a heavy-fermion state that acts as the filter. We determine that the optimal location of the heavy state and hence the onset of the filtering process is at the Dirac point. Our study opens a new avenue for attaining ultrahigh power factors via engineering the EPI in graphene-like semimetals or identifying new compounds that intrinsically possess the featured electronic structure.

\end{abstract}

\maketitle

 
Up to fifty percent of industrial energy consumption is lost as waste heat, which unnecessarily depletes natural resources and negatively impacts the environment. In turn, active cooling is an urgent issue for the increasingly miniaturized and densely packed electronic components.
Waste heat recovery and cooling of high-performance electronics would both benefit from the development of efficient thermoelectric (TE) materials that can directly convert heat into electricity and {\it vice versa}~\cite{Bell1457,Snyder2008,Heeaak9997,BERETTA2018}. Unfortunately, widespread application of TE technologies is hampered by low efficiency and high cost per unit of useful power. A key component of TE efficiency lies in a material's power factor (PF=$S^2\sigma$), which depends on the Seebeck coefficient ($S$) and electrical conductivity ($\sigma$)~\cite{1959PhT}. The main difficulty in achieving high PF is caused by the fact that $S$ and $\sigma$ are negatively correlated with respect to conventional design parameters, such as carrier concentration. A range of creative physical principles have been proposed to improve the PF~\cite{Snyder2008,Arash2015}, such as quantum confinement~\cite{Hicks1993}, resonant levels~\cite{Heremans2008,Heremans2012}, electron energy filtering~\cite{Shakouri1997}, and the so-called ``band engineering"~\cite{Pei2011}.

Good TEs are usually doped semiconductors because relatively high $S$ is guaranteed by the naturally occurring band gap that filters bipolar transport. On the contrary, semimetals and intermetallics are generally deemed impractical as TEs, the culprit being their high electronic density-of-states (DOS) at the Fermi level, which results in low entropy per carrier and often negligible $S$ due to mutually counterbalancing contributions from electrons above and below the Fermi level. However, because they are so conductive, even a moderate value of $S$ could give a very high PF. Ultrahigh PFs are rare, but have been measured in semimetals. A seminal example is graphene~\cite{Novoselov2005,Geim2007,Zhang2005,Zuev2009,Ghahari2016}, whose exceptional room-temperature PF of 345 $\mu$W cm$^{-1}$ K$^{-2}$ holds the record among all TE materials, as recently demonstrated by Duan \etal\cite{Duan2016}. The lattice thermal conductivity of graphene is too high for efficient waste heat recovery, but it shows promise for use in TE cooling. However, besides regular approaches such as improving sample quality and adjusting carrier concentration~\cite{Wang614,Ishigami2007,Martin2007,Dean2010,Duan2016}, there is hardly any other effective way to further improve $S$ while maintaining high $\sigma$ in graphene~\cite{Tabitha2018}. A general approach to enhance $S$ in graphene-like semimetals via is highly desired.

Electron-phonon interaction (EPI) is arguably the most important intrinsic factor that shapes the PF because it usually plays the key role in scattering charge carriers and decreasing $\sigma$~\cite{ziman1960electrons}. Compounds with intrinsically weak EPI can often achieve high PFs: typical examples are  chalcogenides~\cite{SootsmanPbTe,Pei2011,Wang9705} and Heusler compounds~\cite{Fu2015,He13576,Zhoujiawei2018,ba2biau}. Few attempts have been made to exploit EPI to enhance $S$~\footnote{The benefits of EPI through enhancement of $S$ via the phonon-drag effect is a rare exception where EPI arguably plays a beneficial role for the power factor~\cite{Wu1996,Zhou14777}. However, the out-of-equilibrium phonon transport is only achievable at relatively low temperature and also results in high lattice thermal conductivity, both of which limit the practical usefulness of this phenomenon for power generation.}. From a theoretical stand-point, the impact of EPI on $S$ is mediated through the energy-dependent carrier lifetime $\tau(E)$, as seen from the Mott formula~\cite{mott1971} in the degenerate limit:
\begin{equation}
\label{eq:Mott}
\begin{split}
S & = - \frac{\pi^2 k_{\text{B}}^2 T}{3e} \left. \frac{d \ln \sigma(E)}{dE} \right\rvert_{E=E_{\text{F}}}
\end{split}
\end{equation}
Here, $\sigma(E) = n(E) v^2(E) \tau(E)e$ is the energy-dependent electrical conductivity~\cite{Heremans2008}, where $e$, $n(E)$ and $v^2(E)$ are the absolute charge of an electron, electronic DOS, and  the group velocity, respectively. Eq.~(\ref{eq:Mott}) suggests that, even when there is bipolar transport (as occurs in compounds with large $n(E)$ near or through the Fermi level), a large energy dependence of $\tau(E)$ can in principle lead to high $S$. Despite its rarity, evidence for such an effect has been discovered in a recent first-principles study of the intermetallic $B20$-type CoSi~\cite{Xia2019PRApplied}, where an anomalously large intrinsic $S$ arises from strongly energy-dependent lifetimes controlled by EPI. This energy dependence occurs due to a symmetry-mandated intersection between a heavy-effective-mass band and two Dirac-like bands (spin-1 chiral fermions)~\cite{CoSiTP,Chang2017,Takane2019,Rao2019arXiv}, which selectively increases the DOS and the phase space for electron-phonon scattering on one side of the Fermi level. Despite the theoretical interest of this phenomenon in CoSi, however, the position of the heavy band relative to the Dirac point and its effective mass are both fixed by the space group symmetry~\cite{CoSiTP,Chang2017} and cannot be altered in bulk samples, which significantly impedes general applicability of such strategy to existing high-performance thermoelectrics. In light of this discovery, it is natural to ask if such a strategy can be applied to 2D semimetals such as graphene to engineer even higher PFs? Furthermore, what type of band structure would be the best platform for promoting EPI-induced energy filtering? And how may the enhancement in $S$ through EPI-induced energy filtering be optimized against concomitant reduction in $\sigma$ due to EPIs?

We posit that Dirac bands that intersect a flat band at the Fermi level comprise an ideal platform for promoting $S$ and increasing the PF via energy filtering. First, high $\sigma$ is naturally guaranteed due to large Fermi velocities and long carrier lifetimes in linearly dispersing bands~\cite{Park2014,Tabitha2018}. Second, a Dirac cone by itself has zero DOS at the Dirac point and small DOS around it which, if accompanied by a heavier band, would allow the latter to create the sharpest rise in the DOS. Therefore, the derivative $d\tau(E)/dE$ is expected to be most pronounced with Dirac bands in the background, if one assumes that 1/$\tau(E)\propto$~DOS. Moreover, because the DOS of a generic parabolic band in a 2D system exhibits a step function due to quantum confinement, an even stronger energy filtering is expected in 2D than in 3D~\cite{Hicks1993}. To demonstrate this concept theoretically, we select a semimetallic allotrope of monolayer MoS$_{2}$~\cite{Weifeng2014}, whose electronic structure contains the desired crossing between two Dirac bands and a heavy-mass band at the Fermi level~\cite{Weifeng2014}. By means of first-principles calculations with explicit treatment of EPIs, we calculate the PF and confirm its critical enhancement owing to the existence of the heavy band. We further identify that the optimal energy location of the heavy band that maximizes the enhancement of PF is at the Fermi level itself where the Dirac point forms. Potential pathways to realize such a strategy in other systems are also discussed.


Recent development of techniques for EPI~\cite{Giustino2007,NOFFSINGER20102140,Sjakste2015,Verdi2015,Ponce2016,Giustino2017} has enabled first-principles calculations of electrical transport at the mode level~\cite{wuli2015,Qiu2015,Liao2015,jjz2016,Tehuan2017,Sohier2018,Samuel2018,Macheda2018,Jinlong2018}. To compute $S$ and $\sigma$, we assume that carriers are majorly scattered by phonons, the lifetime ($\tau_{i, \mathbf{k}}$) of electronic state with band and wavevector indexed by $i$ and $\mathbf{k}$ can be determined from the imaginary part of the electron self-energy Im($\Sigma$),
	\begin{equation}\label{eq:lifetime}
	\begin{split}
	\frac{1}{\tau_{i, \mathbf{k}}} & =  \frac{2\text{Im}(\Sigma)}{\hbar} =  \frac{2\pi}{\hbar}\sum_{j, \nu} \int_\text{BZ} \frac{d\mathbf{q}}{\Omega_\text{BZ}} \left| g_{ji,\nu}\left(\mathbf{k}, \mathbf{q}\right) \right|^2   \\
	& \times [ \left(n_{\nu, \mathbf{q}}+f_{j, \mathbf{k+q}}\right)  \delta\left( \epsilon_{i, \mathbf{k}}+\omega_{\nu, \mathbf{q}}-\epsilon_{j, \mathbf{k+q}} \right)   \\
	& +  \left(1+n_{\nu, \mathbf{q}}-f_{j, \mathbf{k+q}}\right) \delta\left( \epsilon_{i, \mathbf{k}}-\omega_{\nu, \mathbf{q}} -\epsilon_{j, \mathbf{k+q}} \right) ],
	\end{split}
	\end{equation}
where $\nu$, $\mathbf{q}$, $n$, $\omega$ and $g_{ji,\nu}\left(\mathbf{k}, \mathbf{q}\right)$ are phonon mode index, wave vector, population, frequency, and EPI matrix element, respectively. We calculated Im($\Sigma$) by performing density-functional theory (DFT) calculations~\cite{DFT,QE,DFPT} within the Perdew-Burke-Ernzerhof (PBE) parametrizations of the exchange-correlation functional~\cite{PBE,Hamann2013}. Afterwards, $S$ and $\sigma$  were modeled using the semiclassical Boltzmann transport equation (BTE) under the relaxation time approximation (RTA)~\cite{ziman1960electrons,Madsen2006}. (see Supplemental Material~\cite{SM} for the details of the calculations~\cite{Marzari1997,Souza2001,Marzari2012,Mostofi2008})


\begin{figure}[h]
	\setlength{\unitlength}{1cm}
	\includegraphics[width=1.0\columnwidth,angle=0]{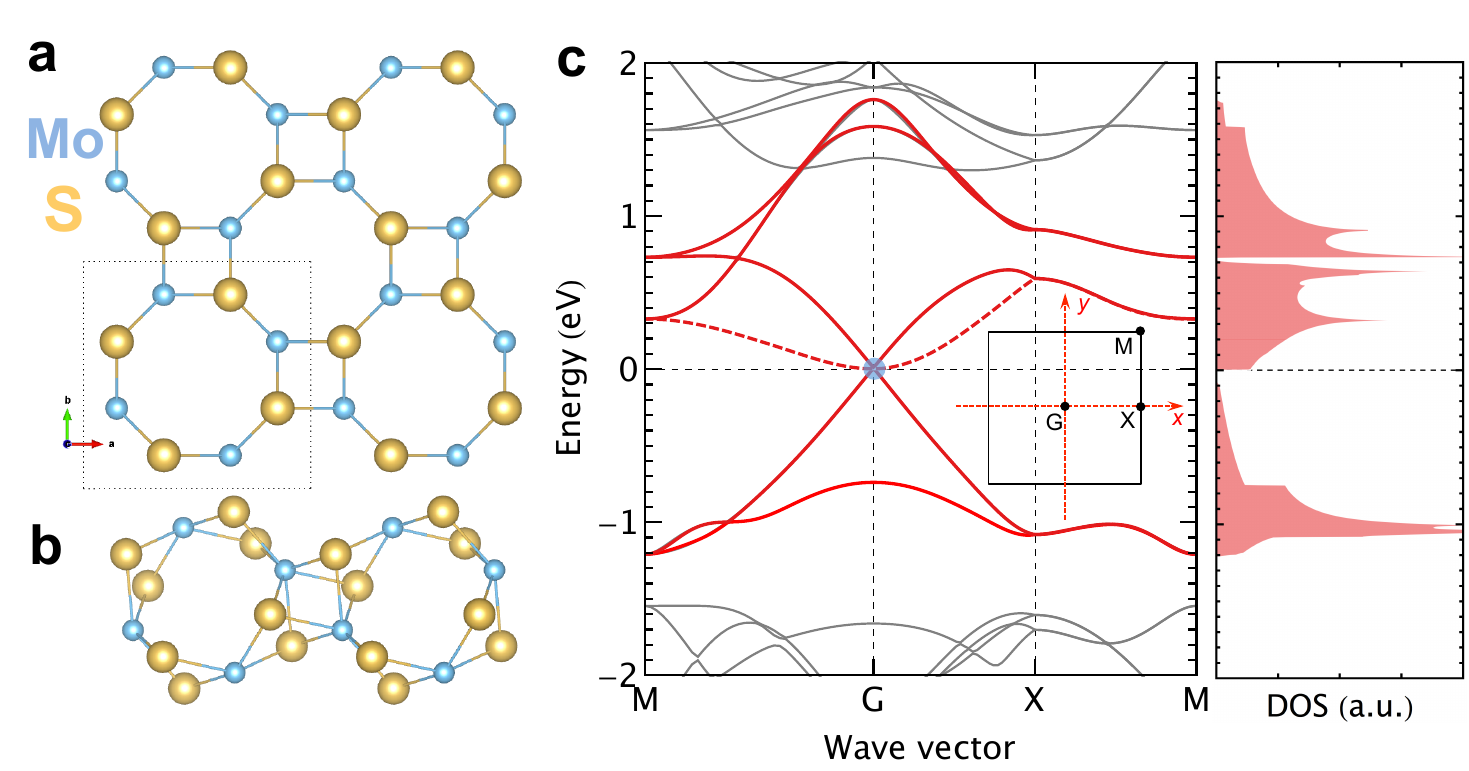}
	\caption{(a) The top view and (b) the side view of so-MoS$_{2}$. (c) The electronic band structures of so-MoS$_{2}$ and the total density of states. The inset in (c) shows the Brillouin zone and the high symmetry points. The Fermi level (horizontal dashed black lines) at 0 K is shifted to 0 eV with the Dirac point denoted by the blue disk. The Wannier states that expand the manifold, composed of six bands near the Fermi level, are colored in red, wherein the heavy band responsible for energy filtering is highlighted by the dashed curve.}
	\label{fig:StrBand}
\end{figure}

The monolayer MoS$_{2}$ allotrope is constructed of repeating square-octagon pairs in a square lattice (abbreviated as so-MoS$_{2}$)~\cite{Weifeng2014}, as shown in Fig.~\ref{fig:StrBand}(a) and (b). The resulting structure bears the $P4$ symmetry, containing four Mo and eight S atoms within the primitive cell. The band structure of so-MoS$_{2}$ features a Dirac point at the Fermi level, as shown in Fig.~\ref{fig:StrBand}(c), where the Dirac bands are crossed by a heavy band (HB) that is nearly parabolic. The Dirac bands and the heavy band have predominantly Mo $d_{x^2-y^2}$ and $d_{z^2}$ character, respectively (see Ref.~[\onlinecite{Weifeng2014}]), and the near-crossing between them is accidental (i.e., not mandated by crystal symmetry). The DOS exhibits a sharp Heaviside-like onset at the Fermi level due to the quantum-confined 2D parabolic band, and also a linear increase with energy that is characteristic of the linearly dispersing Dirac bands~\cite{Hicks1993}.

\begin{figure}[htp]
	\setlength{\unitlength}{1cm}
	\includegraphics[width=1.0\columnwidth,angle=0]{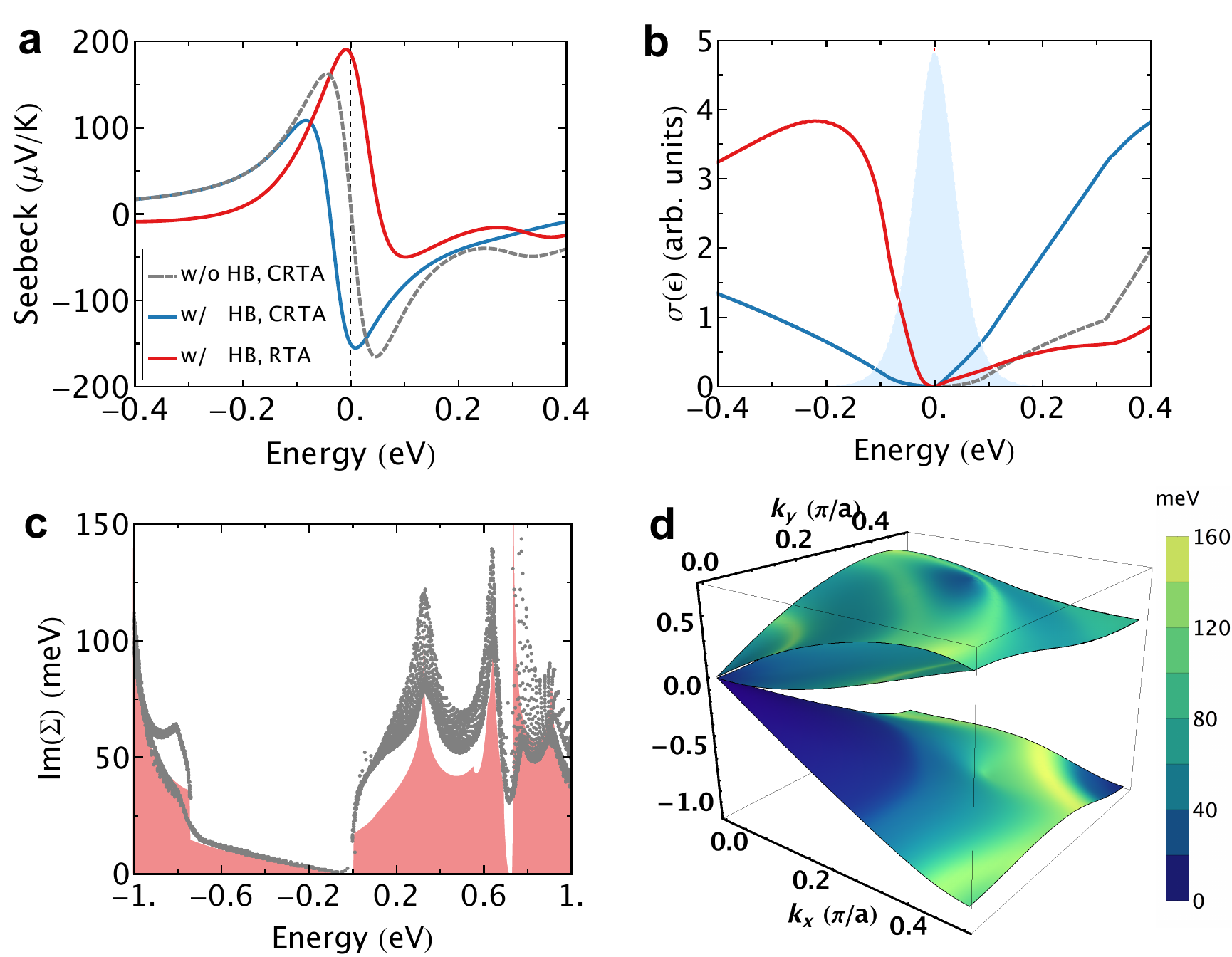}
	\caption{(a) Calculated Seebeck coefficient at 300 K as a function of chemical potential using band structures with (solid blue lines) and without (dashed gray lines) the heavy band (HB) under the constant relaxation time approximation (CRTA). The red line traces the values calculated with the HB under relaxation time approximation (RTA). (b) The same as (a) but for energy-dependent electrical conductivity $\sigma(E)$. The region shaded off-blue is the energy window that contributes to charge transport. (c) The imaginary part of the electron self-energy (gray dots) near the Dirac point at 300 K, which scales well with the total density of states (shaded red). (d) Three dimensional contour plot of momentum- and energy-resolved band structure, color-coded with the magnitude of the imaginary part of the electron self-energy.}
	\label{fig:Seebeck}
\end{figure}

Our investigation of the electronic transport properties is focused on the role played by the HB in determining the transport behavior of the mobile Dirac states. First, by assuming the constant relaxation time approximation (CRTA), we examine the effect of the HB's DOS and group velocities, the static features of a band, on $S$. We compare the cases where $S$ is calculated with and without the HB, and for the sake of fair comparison, we assume that the Fermi level does not shift in the absence of the HB. As shown in Fig.~\ref{fig:Seebeck}(a), the calculated $S$ without the HB behaves symmetrically about the Dirac point, reflecting the symmetry of the Dirac cone. Upon the introduction of the HB, $S$ experiences a significant negative shift at the Dirac point. This complies with the fact that the HB provides an excess electron and drives the system to be $n$-type population-wise. Finally, upon the introduction of EPI lifetimes ($\tau_{i, \mathbf{k}}$) the $S$ profile undergoes a drastic sign reversal from -153~$\mu$V/K to 186~$\mu$V/K at the Dirac point, indicating that holes dominate transport by attaining longer lifetimes than electrons in spite of the relative deficiency in population. This is a far greater lifetime-induced sign-reversal than in the aforementioned $B20$-type CoSi (14~$\mu$V/K to -68~$\mu$V/K)~\cite{Xia2019PRApplied}. Also, there exists an energy window near the Dirac point where $S$ retains larger values than if there were only the Dirac bands present, indicating the possibility to achieve higher PF in the presence of the HB. This is in a striking contrast to the results under the CRTA, which is biased toward high DOS and incorrectly predicts higher PF for the Dirac-and-heavy composite bands. Our findings highlight the necessity of examining the lifetime effects for reliably capturing the PF enhancement through $S$ by means of sharp DOS~\cite{Heremans2002,Mahanti2006,Grossman2010,Parker2013,Bilc2015}.

The impact of the HB on $S$ can be readily understood by examining the energy dependence of $\sigma(E)$ shown in Fig.~\ref{fig:Seebeck}(b). CRTA incorrectly predicts higher $\sigma(E)$ for electrons simply because of their excess population over that of the holes. RTA, however, reverses this trend, and the holes have significantly higher $\sigma(E)$ due to much longer lifetimes. Eq.~(\ref{eq:Mott}) shows that the sign of $S$ is determined by the energy-derivative of $\sigma(E)$, which within the RTA is ultimately traced back to the energy dependent $\tau(E)$ shown in Fig.~\ref{fig:Seebeck}(c). The strong enhancement of Im($\Sigma$) above the Fermi level indicates a heavily preferential scattering of electrons. The momentum-dependent Im($\Sigma$) profile displayed in Fig.~\ref{fig:Seebeck}(d) further clarifies that electrons are indeed strongly filtered via EPI throughout the Brillouin zone. The onset of the enhancement in Im($\Sigma$) and $\sigma(E)$ is clearly correlated with the sharp increase in the DOS at the Fermi level, as electron scattering rates are proportional to the availability of the final states that conserve energy and momentum (the EPI phase space), i.e, 1/$\tau\propto$~DOS, as implied by Eq.~(\ref{eq:lifetime}).

Our analysis shows that the drastic EPI-induced change in Im($\Sigma$) and $\sigma(E)$ across the Fermi level opens up an effective band gap, which allows the emergence of ``majority carriers" not via excess population, but via more mobile transport. Though it is not the typical energy gap of semiconductors, it can be just as effective in resisting bipolar transport and sustaining high $S$. The reason for the more prominent filtering effect in the 2D so-MoS$_{2}$ compared to that in the 3D $B20$-type CoSi is the quantum-confined DOS of the former. It is also worth noting that such filtering effect relying on intrinsic EPI at ambient conditions is particularly appealing, especially compared to the recently proposed energy filtering by edge-bulk interactions in topological insulators~\cite{Xu2014SB} that relies on quantum effects at cryogenic temperatures.

The observation that the peak of $S$ occurs at the sharp onset of the DOS invites the question whether higher $S$ and PF may be accessible if the HB were shifted in energy relative to the Dirac point ($\Delta E$). Two limiting cases are of particular interest: (i) when the HB strongly overlaps with the Dirac bands and (ii) when the HB is far away from the Dirac point. We perform a set of numerical experiments by manually shifting the HB by $\Delta E$ = $\pm$0.2, $\pm$0.1, 0.0 eV, as shown in Fig. S3(a) in the Supplemental Material~\cite{SM}. For a fair comparison of the HB-effect on the transport of holes and electrons, we assume that the Fermi level occurs at the Dirac point regardless of the shift. Meanwhile, we keep the EPI matrix elements fixed to their original values and only change the phase space according to the shifted eigenenergies. This treatment draws merit from our earlier observation that Im($\Sigma$) is dominated by the EPI phase space. The sharp onset in Im($\Sigma$) is consistently correlated with the rapid variation of DOS regardless of $\Delta E$ (see Fig. S3(b)-(f) in the Supplemental Material~\cite{SM}), indicating that the phase space rather than individual EPI matrix elements is the dominant factor. This allows us to realistically model the dependence of $S$, $\sigma$, and PF on the HB position $\Delta E$.

\begin{figure}[h]
	\setlength{\unitlength}{1cm}
	\includegraphics[width=1.0\columnwidth,angle=0]{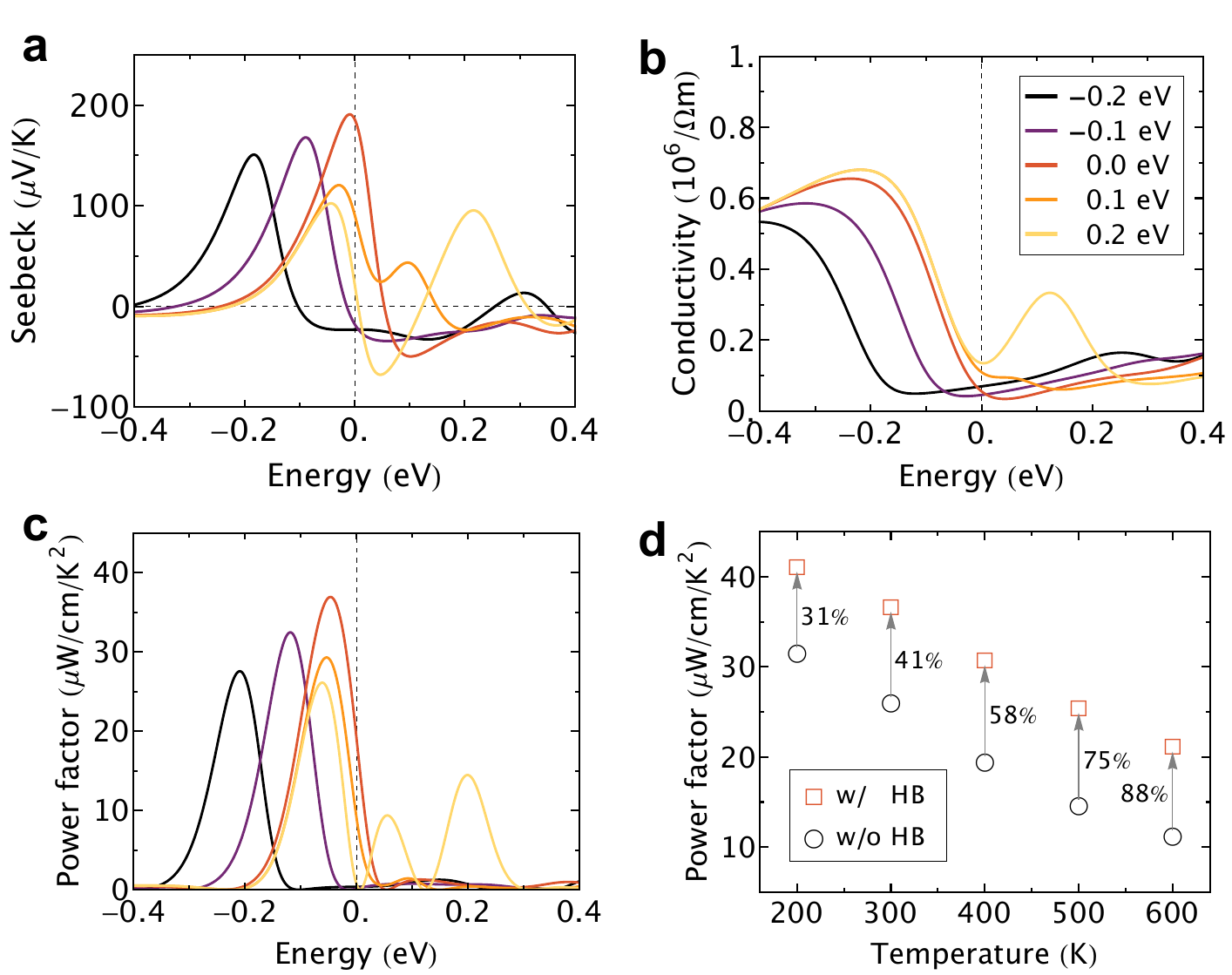}
	\caption{Calculated (a) Seebeck coefficient, (b) electrical conductivity, and (c) the power factor as functions of the chemical potential at 300 K under the relaxation time approximation. The lines are colored to indicate differential energy shifts ($\Delta E$ = $\pm$0.2, $\pm$0.1, 0.0~eV) imposed on the heavy band (HB). (d) Comparison of the maximum power factor without (black circles) and with (red squares) the HB at the optimal chemical potential, namely, at the Dirac point, from 200 to 600~K. The numbers attached to the vertical arrows indicate the relative enhancement in the power factor.}
	\label{fig:sbpf}
\end{figure}

With the HB shifted up by $\Delta E$ = 0.1~eV, an additional peak in $S$ appears just below the Fermi level, signifying the contribution from the low-lying Dirac bands. When the HB is shifted further away, by $\Delta E$=0.2~eV, the magnitude of $S$ below the Fermi level becomes saturated, corresponding to the limiting case (ii).  As for $\sigma$, it is strongly suppressed for electrons when the HB is shifted up, resulting in comparatively higher $\sigma$ for holes [see Fig.~\ref{fig:sbpf}(b)]. Hole conductivity, on the other hand, increases as the HB moves away from the Dirac cone, and attains saturation when $\Delta E >$ 0.0~eV. As a consequence of compromise between $S$ and $\sigma$, all peak PF values are associated with positive $S$, reflecting hole-dominated transport arising from purely Dirac bands. Overall, it is evidently important that the HB is not shifted down in energy, which suppresses $\sigma$ for holes and also damages $S$ and the PF due to undesirable hole scattering. Shifting the HB up in energy suppresses the PF towards the lower bound of 26 $\mu$W cm$^{-1}$ K$^{-2}$ due to insufficient filtering of electrons.

The largest enhancement in both $S$ and the PF are found for $\Delta E$ = 0.0~eV, the un-shifted original case where the band crossing occurs at the Dirac point. Specifically, the largest PF peak of 37 $\mu$W cm$^{-1}$ K$^{-2}$ is predicted when the HB crosses the Dirac bands at the Dirac point. The results unambiguously show that, compared to the case of bare Dirac bands, an approximately 41\% enhancement of PF at 300~K is achieved if a HB exists at an optimal energy, owing to the asymmetric $\sigma$ and energy-filtering-enhanced $S$. The relative PF enhancement is even more pronounced at higher temperature, i.e., from 31\% to 88\% when temperature is increased from 200~K to 600~K, as shown in Fig.~\ref{fig:sbpf}(d). We emphasize that the \textit{absolute} value of the PF depends somewhat on the specific approach for solving the Boltzmann transport equation \cite{wuli2015,Ponce2016,Jinlong2018,Macheda2018}. However, the main concept advanced in this work is the \textit{relative} enhancement of the PF due to the coexistence of Dirac cone and a parabolic band, which is expected to be universal and largely independent of the adopted approximation~\footnote{For example, we have found that the relative enhancement of the power factor remains as high as 40\% at 300~K when carrier lifetimes are more accurately estimated by accounting for the velocity factors~\cite{wuli2015}, even though the absolute values of power factor decrease.}.

The principle of PF enhancement unveiled here via phase-space modulation should be applicable to the engineering of higher PFs in 2D materials with graphene-like Dirac bands. Furthermore, the lower dimensionality and the accidental (i.e., not mandated by  symmetry) near-degeneracy of heavy and Dirac bands allows extra tunability to achieve maximum PF, for example, by adjusting the effective mass of the parabolic band via epitaxial stress. Concerning practical realization, the PF enhancement in graphene through EPI-induced energy filtering may be achieved by forming heterostructures under gate-voltages~\cite{Duan2016}, where upon a proper choice of the semiconducting substrate, the VBM or CBM of the substrate could cross the Dirac point~\cite{LiX2013}. Compounds that intrinsically possess the desired band structure also merit pursuit. Recent studies show that a coexistence of Dirac and flat bands in fact occurs in the so-called Kagome lattices, which consist of equilateral triangles and hexagons~\cite{ZHONG201665,Syozi1951}. In particular, the double Kagome lattice show promise to possess the featured heavy-and-Dirac crossing~\cite{Huang2018}. Discovering and/or realizing these types of semimetallic compounds as highly efficient TEs is a promising step forward.

We have shown that engineering of the electron-phonon interaction in semimetals can significantly enhance their Seebeck coefficient and thermoelectric power factor. In particular, we find that enhanced electron-phonon scattering in the vicinity of a sharp spike in the electronic DOS results in strong energy filtering of electrons and yields high power factors. Since electron-phonon interaction is intrinsic to the material and largely independent of processing and composition, this represents an attractive design strategy. Our theoretical proposal, using a two-dimensional gapless MoS$_{2}$ allotrope as the case study, is a proof-of-concept that the crossing of heavy bands and massless Dirac bands at the Fermi level is a favorable type of band structure for combining high $S$ and $\sigma$. Our finding not only opens a new avenue for further power factor enhancement in compounds with graphene-like band structures, but also provides a guidance for identifying previously unknown semimetallic and metallic compounds with high thermoelectric performance.


\begin{acknowledgments}
\textbf{Acknowledgments: } The research was conceived and designed by Y.X., V.O. and C.W.. DFT calculations of thermoelectric properties were conducted by Y.X. All authors provided critical feedback and helped shape the research, analysis and manuscript. Y.X. and C.W. acknowledge support from the U.S. Department of Energy under Contract No. DE-SC0015106. J.P. and V.O. acknowledges financial support from the National Science Foundation Grant DMR-1611507. This research used resources of the National Energy Research Scientific Computing Center, a DOE Office of Science User Facility supported by the Office of Science of the U.S. Department of Energy under Contract No. DE-AC02-05CH11231.
\end{acknowledgments}

\bibliography{MoS2}
\end{document}